\documentclass[final, 12pt]{elsarticle}

\usepackage{graphicx}
\usepackage{epsfig}
\usepackage{listings}
\usepackage{hyperref}
\usepackage{verbatim}
\usepackage{amssymb}
\usepackage{makecell}
\usepackage[format=plain, belowskip=-20pt,aboveskip=2pt,labelfont=bf,textfont=it]{caption}

\usepackage{lineno}
\newcommand{\listingsfont}{\ttfamily}
\usepackage{etoolbox}
\patchcmd{\pprintMaketitle}
 {\ifvoid\absbox\else\unvbox\absbox\par\vskip10pt\fi}
 {\ifvoid\absbox\else\clearpage\unvbox\absbox\par\vskip30pt\fi}
 {}{}
\patchcmd{\pprintMaketitle}
 {\hrule\vskip12pt}
 {}
 {}{}
\patchcmd{\pprintMaketitle}
 {\hrule\vskip12pt}
 {}
 {}{}
\appto{\pprintMaketitle}{\clearpage}

\usepackage{etoolbox}
\makeatletter
\patchcmd{\ps@pprintTitle}
  {Preprint submitted}
  {Accepted manuscript}
  {}{}
\makeatother

\journal{Journal of Parallel and Distributed Computing}

\begin{document}

\begin{frontmatter}


\title{High level programming abstractions for leveraging hierarchical memories with micro-core architectures}



\author[epcc]{Maurice Jamieson}
\ead{maurice.jamieson@ed.ac.uk}
\author[epcc]{Nick Brown}
\ead{n.brown@epcc.ed.ac.uk}

\address[epcc]{EPCC, The Bayes Centre, 47 Potterrow, Edinburgh}

\begin{abstract}
Micro-core architectures combine many low memory, low power computing cores together in a single package. These are attractive for use as accelerators but due to limited on-chip memory and multiple levels of memory hierarchy, the way in which programmers offload kernels needs to be carefully considered. In this paper we use Python as a vehicle for exploring the semantics and abstractions of higher level programming languages to support the offloading of computational kernels to these devices. By moving to a pass by reference model, along with leveraging memory kinds, we demonstrate the ability to easily and efficiently take advantage of multiple levels in the memory hierarchy, even ones that are not directly accessible to the micro-cores. Using a machine learning benchmark, we perform experiments on both Epiphany-III and MicroBlaze based micro-cores, demonstrating the ability to compute with data sets of arbitrarily large size. To provide context of our results, we explore the performance and power efficiency of these technologies, demonstrating that whilst these two micro-core technologies are competitive within their own embedded class of hardware, there is still a way to go to reach HPC class GPUs.
\end{abstract}

\begin{keyword}
Parallel programming languages ; Interpreters ; Runtime environments ; Hardware accelerators ; Neural networks
\end{keyword}

\end{frontmatter}

\section{Introduction}

Micro-core architectures combine many simple, low power, cores on a single processor package. Their low power and low cost makes them attractive for multiple domains and we are seeing the embedded and HPC worlds converging. The embedded world which has always focused on power efficiency is now interested in parallelism, and the HPC community having to consider power efficiency in order to facilitate realistic future exa-scale machines. These micro-core architectures, providing significant parallelism and performance for low power are therefore of great interest to both communities and have been at the heart of the top machine in the Green 500 until March 2019 \cite{green500}.

Very often machines built around micro-core architectures exhibit multiple levels of memory hierarchy, from the small and fast on-core scratch pad memory expanding out to slower but larger memory spaces. Knowing where about to place their data in the memory hierarchy and then retrospectively changing this if it is not optimal adds significantly to the burden placed upon the programmer. This problem is magnified by micro-core architectures due to the immaturity of programming tools and the fact that the hierarchy is often deep and memory spaces comprise of KBs of manually controlled memory close to the core rather than automatic caches of many MBs. Hence, with micro-cores, not only does the programmer need to correctly control data placement for performance, but they also need to get this right for their code to even run in the first place.

The severely constrained nature of micro-cores makes the challenge of data placement and transfer a difficult one. In this paper we ues Python as a vehicle for presenting and demonstrating our abstractions for offloading kernels to micro-core accelerators such that the programmer can process arbitrarily large data sets on micro-cores and control data placement in the memory hierarchies without having to deal with the low level, complex, nitty-gritty details of how data is physically moved. In short the contributions of this paper are:

\begin{itemize}
    \item Demonstration that a pass by reference model, similar to CUDA's unified virtual addressing, is mandatory for enabling micro-core to process arbitrarily large data-sets. We explore the performance characteristics of this approach and role that pre-fetching plays in optimising data transfer.
    \item Demonstration that, for micro-cores, memory kinds enable the programmer to concisely express where in the memory hierarchy their data is located, with the runtime and kinds themselves then responsible for low level data transfer.
    \item A general performance and power efficiency comparison of micro-cores against embedded and HPC class hardware technologies.
\end{itemize}

This paper is laid out as follows, after describing general background, specifics of the hardware used and related work in Section \ref{sec:bg}, Section \ref{sec:directives} then focuses on the abstractions we have developed to enable the programmer to seamlessly leverage memory hierarchies in their code. Section \ref{sec:implementation} then discusses some of the implementation challenges that had to be addressed to adopt these new extensions. In Section \ref{sec:results} we use a machine learning code for detecting lung cancer in 3D CT scans as a benchmark, run on both the Epiphany-III and MicroBlaze based micro-cores. These experiments are used to illustrate the performance of our approach and general power efficiency of micro-cores, comparing that against characteristics of other common technologies. We then draw conclusions in Section \ref{sec:conclusions} and discuss further work.

\section{Background and related work}
\label{sec:bg}
There are numerous micro-core architectures such as the PEZY-SC2 \cite{pezy-sc} which powered the top Green 500 machine until it was decommissioned in March 2019, although at the time of writing at 17.6 GFLOPS/Watt is still more energy efficient than the current number one GPU-based machine \cite{green500}. The Kalray Boston \cite{kalray}, the Celerity \cite{ajayi2017celerity}, and numerous soft cores are other examples of micro-cores, and these technologies are at varying levels of availability, maturity and cost. The work and experiments described in this paper focuses on two very different types of micro-core, the Epiphany \cite{epiphany-intro} and MicroBlaze \cite{microblaze}. The Epiphany is arguably one of the most ubiquitous of these micro-cores, developed by Adapteva and packaged as a single physical chip comprising of low power cores. On the Epiphany-III each of these cores consists of a RISC CPU, 32KB high bandwidth on-core local memory, DMA engine and network interface. Whereas the Epiphany is a physical chip, Xilinx's MicroBlaze is instead a semiconductor intellectual property core, known as an IP block, and used in conjunction with interconnection IP blocks, to configure a Field Programmable Gate Array (FPGA) to present itself as a multi-core MicroBlaze CPU. Known as a soft-core, from the end programmer's perspective this chip looks like a CPU, but crucially this approach is much cheaper than physical cores as there is no need for expensive manufacturing, and significantly more flexibility in configuration than a physical CPU. Out of the numerous soft-cores available, the MicroBlaze is amongst the most ubiquitous, not least because it is developed by Xilinx, arguably the world leading FPGA vendor. Irrespective of whether the implementation is a physical or soft CPU, these technologies contain many cores, each with very limited amounts of memory, and the reason for picking these two technologies in our experiments is both their ubiquity, and also representation of a specific class of micro-cores.

The micro-core architecture is applicable to a wide range of problem domains and performance levels close to 2 GFLOPs per core have been demonstrated \cite{epiphany} in the field of signal processing on the Epiphany chip. The major advantage of this technology is the power efficiency, for instance the most common Epiphany is the 16 core version 3 (Epiphany-III), manufactured at a process size of 65nm, delivers 32 GFLOPs and draws a maximum of 2 Watts (16 GFLOPs/Watt.) There have been studies comparing the performance and power efficiency benefits of FPGAs against GPUs \cite{fpgavsgpu} and the Zynq-7020 (28nm process size) used in this paper has a theoretical peak performance of 180 GFLOPs and 72 GFLOPs/Watt \cite{fpgavsgpu}. Specifications regarding the MicroBlaze are more difficult because it also depends on the physical FPGA that is being used, although it has been claimed that soft-cores retain many of the power efficiency benefits of FPGAs \cite{castells2016energy}.

In addition to the micro-core, one also requires a board to mount this chip and expose it to the outside world. In this paper we use two such boards, one for each technology. For the Epiphany, the same company also developed the Parallella \cite{parallella} single board computer (SBC). This machine combines a host dual core ARM A9 CPU, with 1 GB of RAM and the 16 core Epiphany-III. Due to limitations in the Parallella, whilst the theoretical off-chip bandwidth of the Epiphany III is 600 MB/s, the maximum obtainable in practice is 150 MB/s \cite{castro2018energy}. For MicroBlaze experiments we use the Pynq-II SBC, mounting a Xilinx Zynx-7020 and 512 MB RAM. Zynq-7020 FPGAs, with an off-chip bandwidth of 131.25 MB/s, contains both a dual core ARM A9 CPU and re-configurable FPGA fabric on the same physical package. This specific FPGA comprises of 53,200 programmable Look-Up Tables (LUTs), and around 627 KBs of block RAM (BRAM) \cite{zynq}. This means that we can fit a maximum of eight 64KB MicroBlaze CPUs, and supporting infrastructure IP blocks, onto the Zynq, which is the configuration used throughout this paper. Whilst we have picked these technologies due to their availability and popularity, in our opinion the MicroBlaze is the more interesting target due to the significant commitment by Xilinx, and active development of many micro-core style soft-cores, including implementations of the RISC-V architecture \cite{celio2017boomv2} \cite{picorv32}.


The programming of these micro-cores is technically challenging, with both technologies supporting C via the GCC tool chain. Whilst some approaches beyond using C with the low level hardware specific library, such as OpenCL \cite{cprthr}, BSP \cite{ebsp}, OpenMP \cite{ompi} and MPI \cite{epiphany-mpi} have been developed, these are at different levels of maturity and still require the programmer to explicitly program the chip using C at a very low level. Indeed, Xilinx's Pynq-II board has been designed around ease of use, loading up a default configuration of three MicroBlaze cores, and presenting a Python interface via the Jupyter notebook. However, Python only runs on the host ARM CPU of the Pynq-II and the programmer must still write C code to execute directly on each MicroBlaze and interface it appropriately with the host code.

\begin{figure}
	\centering
	\includegraphics[width=0.80\textwidth]{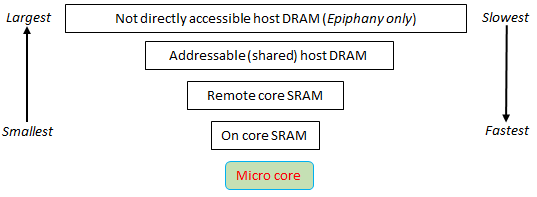}
	\caption{Epiphany and MicroBlaze memory hierarchy}
	\label{fig:memhierarch}
\end{figure}

This programmability challenge is made more severe when one considers the tiny 32KB of memory per core on the Epiphany and 64KB on the MicroBlaze. Whilst some of the board's main memory is directly addressable by the micro-cores, there is a significant performance penalty in accessing this and programmers have to either keep their programs and data within the micro-core memory limits or design their codes to pre-fetch for reasonable performance. Regardless, this adds considerable additional complexity to any non-trivial codes. Figure \ref{fig:memhierarch} illustrates the memory hierarchy for both the Epiphany-III running on the Parallella and multi-core MicroBlaze running on the Pynq-II. The only difference between the two is that the Epiphany/Parallela combination contains a top-level that is not directly accessible to the micro-core whereas the main memory of the MicroBlaze/Pynq-II is all directly accessible by the MicroBlaze cores.

\subsection{Existing accelerators offload approaches}
\label{sec:otheroffload}
There are numerous offloading approaches that target accelerators in general, and specifically GPUs, but none of these technologies currently support micro-cores due to the memory limit. CUDA \cite{nickolls2008scalable} and OpenCL \cite{stone2010opencl} are arguably the most popular approaches, and OpenCL even supports FPGAs \cite{muslim2017efficient}. With both these technologies, the programmer decorates specific kernels in their code to run on the device and they also have available to them a host based library of functionality which enables interaction with the device. The sort of functionality supported from the host is the copying of data, activation of specific kernels, or queuing up of kernels. 

CUDA has supported Unified Virtual Addressing (UVA) since CUDA 4, which enables the programmer to view GPU and CPU memory as one large address space. If pointers on the GPU point to some memory held on the CPU then transfer happens transparently, although there might be some performance penalty. This corresponds to the \emph{Addressable external DRAM} level and below in the memory hierarchy of Figure \ref{fig:memhierarch}. A limitation of UVA is that the implementation of this is fairly basic and if the GPU is repeatedly accessing CPU memory, then a significant performance penalty could be incurred. Pascal architectures and later, supported by CUDA 6, implement Unified Memory (UM). This provides UVA's virtual address space and additional support to transparently move memory to the GPU, treating the GPU memory more like a large cache. The simplest memory movement strategy moves memory on a page fault, but there are more advanced functionalities such as pre-fetching to minimise the overhead of data migration. In experimentation \cite{umstreaming} it has been shown that UVA's explicit copying of memory to a pre-allocated buffer on the GPU is the still fastest, but this is comparable with the pre-fetching performance of UM. Both of these approaches are significantly faster than non-prefetched UM, due to the overhead of page migration. However, other experiments of \cite{landaverde2014investigation} report more limited success of UM pre-fetching, especially with irregular data patterns.

An important aspect of UVA and UM is that the hardware must provide support for these technologies which adds extra complexity to the chip, especially UM \cite{agarwal2016selective}. Both the micro-core systems used in this paper support some degree of UVA, which itself required some addition to the micro-core chip and target machine software. For instance, in the case of the MicroBlaze running on the Pynq-II, UVA is facilitated by the existing Zynq IP block at the FPGA level, and Xilinx libraries running on the host ARM. But micro-cores can apply to a variety of different target machines, and as such hardware level UVA can not always be assumed, nor can direct memory access to higher levels of the memory hierarchy in Figure \ref{fig:memhierarch} be guaranteed.

A downside of CUDA and OpenCL is that is can require explicit, and sometimes fairly low level, user code to perform actions such as allocating and data movement. Pragmas are an alternative approach for limiting the amount of host level support code, with the programmer decorating specific parts of their code with directives which then instruct the compiler to extract these as kernels, and execute them on the appropriate device. Common approaches include OpenACC \cite{openacc} and OpenMP 4.0 \cite{openmp}, and, for instance, the \emph{target} directive of OpenMP 4.0 marks a region of code to execute on a device and a mapping between the device and host memory can be specified. The \emph{declare target} directive can be used to declare global variables on the target device and kernels can execute concurrently by wrapping them as an OpenMP \emph{task}. Numba \cite{numba} is an annotation driven approach for offloading Python kernels to GPUs. The programmer decorates specific functions in their code and these will be executed on the GPU and perform all data movement necessary. However, Numba requires on-device memory significantly in excess of that provided by micro-cores and doesn't provide any significant support for hierarchies of memory.

\subsection{ePython}
\label{sec:bg_epython}
ePython \cite{pyhpc} is an implementation of Python, initially developed for the Epiphany, and now ported to other micro-core architectures including the MicroBlaze. The primary purpose of ePython was initially educational, but it is also applicable as a research vehicle for understanding how best to program these architectures and prototyping applications on them. Due to the memory limitations of these architectures, the ePython interpreter (written in C) fits into 24KB of memory, with the remaining memory used for user byte code, the stack, heap and communications. It is possible for byte code, the stack and heap to overflow into shared memory but there is a performance impact of this. ePython also supports a rich set of message passing primitives such as point to point messages, reductions and broadcasts between the cores. 

At 24KB ePython is by far the smallest implementation of Python and specially designed for highly parallel systems. MicroPython \cite{micropython} is another implementation of Python designed for micro controllers but crucially MicroPython is hundreds of KBs, and whilst this is small in comparison to many Python interpreters such as CPython, it is still significantly above the in-core memory limitations of micro-core architectures such as the Epiphany and MicroBlaze. The other big difference between MicroPython and ePython is that of parallelism. Whilst there is multi-threading in MicroPython, the programmer is not able to write distributed memory style parallel codes in MicroPython running over a multiple cores concurrently, which ePython trivially supports.

A major aim of ePython was to allow the programmer to view the micro-cores as an accelerator and offload kernels from a host CPU to these computational \emph{engines}. In addition to being able to execute Python codes directly on the micro-cores, an abstraction for offloading kernels from large scale, existing codes running on the CPU to the micro-cores has been developed \cite{epython-poster}. There are three major components to ePython, the 24KB ePython Virtual Machine (VM) running on each micro-core comprising of an interpreter and runtime, general supporting functionality running as a process on the host CPU, and thirdly an ePython module which is imported into a user's Python code running under any Python interpreter such as CPython on the host. Previously the ePython module running on the host did not directly communicate with the micro-cores and instead communication was marshalled via the host ePython process.




Listing \ref{lst:offloaddirectives} illustrates an example code, run under any Python interpreter such as CPython, on the host. A function, in this case \emph{mykernel}, is decorated with the \emph{offload} directive (located in the ePython module) at line 11. When such an offloaded function is called, such as is the case at line 20, the kernel and associated arguments are transparently transferred onto the micro-cores which will then execute the kernel with associated data on the cores using the ePython VM as an \emph{engine}. Any return values will then be copied back from the micro-core to the users code on the host. In this example two lists of numbers, \emph{nums1} and \emph{nums2} are filled with 1000 random numbers on the host (lines 7 to 9.) These are then copied onto the micro-cores as arguments when the \emph{mykernel} function is invoked at line 20. On the micro-cores each element of the first list is summed with the corresponding element of the second list and the result returned. By default kernel execution is blocking and runs on every micro-core, for instance with the Epiphany-III, sixteen copies of the kernel will all run concurrently with the arguments independently passed to each core. Hence, in this case, sixteen identical results, one from each micro-core, are copied back in a list, each element representing the return value(s) of the kernel executing on the corresponding core. ePython provides numerous options that the programmer can pass to the \emph{offload} directive that will further specialise this kernel, such as running on a subset of cores, running asynchronously and policies of scheduling the kernel.

\begin{lstlisting}[frame=lines,caption={Python offload example for summing two lists of numbers},label={lst:offloaddirectives}]
from epython import offload
import random

nums1=[0] * 1000
nums2=[0] * 1000

for i in range(1000):
  nums1.append(random.randrange(0,100,1))
  nums2.append(random.randrange(0,100,1))

@offload
def mykernel(a, b):
  ret_data=[0] * len(a)
  i=0
  while i < len(a):
    ret_data[i]=a[i] + b[i]
    i+=1
  return ret_data
  
print mykernel(nums1, nums2)
\end{lstlisting}

The problem with the code illustrated in listing \ref{lst:offloaddirectives} is that of memory requirements for lists \emph{a}, \emph{b} and \emph{ret\_data}. Each of these lists is approximately 4KB and the micro-cores are so seriously memory constrained that, combined with the 24KB ePython interpreter and byte code, it is likely one or more of these won't fit in the Epiphany memory, being forced to reside in the much slower main board shared memory. In this case there isn't really much the programmer can do, and the situation becomes more serious when the programmer wishes to process larger amounts of data that do not even fit in the shared part of main memory. This limitation has been a significant issue for the ePython offload approach and one in which, until the work of this paper, meant that only small data sizes could be tackled by the technology.

It is also possible to use a device resident data approach, a technique commonly used with GPUs, where variables are allocated directly on the device and the programmer explicitly controls when values are copied on and off. These values don't then need to be transferred on every kernel invocation, which can be especially useful when kernels are executing multiple times on same data. In ePython a \emph{define\_on\_device} API call is provided which the programmer can call from their host code to allocate a variable on the device, data can then be copied on and off using \emph{copy\_to\_device} and \emph{copy\_from\_device} API calls respectively.

\section{Modifying the offload behaviour for micro-cores}
\label{sec:directives}
In this section we describe the two major aspects of our work that enable micro-cores to run kernels handling arbitrarily large amounts of data. The first is a change to the behaviour of function offloading, where instead of eagerly copying the entirety of argument data to the micro-cores on kernel invocation, a reference to this data is passed to the micro-cores and data retrieved on demand. Our second contribution is the use of memory kinds to control where in the memory hierarchy data should reside and these kinds contain functionality to enable transparent accessed by the programmer.

\subsection{Passing kernel data by reference to micro-cores}
\label{sec:passbyref}
The semantics of Python, and many similar dynamic languages, is pass by reference where the reference, or pointer, of an object is passed to a function rather than the data itself. This is important because, if the programmer modifies the data during function execution, then it is not the function's copy of the data but the original data itself that is modified. However as described in Section \ref{sec:otheroffload}, technologies for offloading kernels to accelerators commonly pass by value instead, explicitly copying the entire kernel data and even UM migrates pages of memory on demand. This eager copying of data, whether it be the entire kernel's data before execution, or a page of data before access, makes a lot of sense for GPUs. However it relies on the assumption that there is sufficient memory available on the accelerator to hold this data, and the kernel or data access can not start until transfer has completed. These two factors, and most critically the memory requirement for holding all the data, are significantly limitations when applied to micro-core architectures. Coupled with the fact that no micro-core architectures support hardware level memory migration, as required with UM, then a programming and/or runtime level solution must be found.

In our approach we have modified the behaviour of kernel invocation such that instead of copying the entire data over to the micro-cores (such as all of \emph{nums1} and \emph{nums2} in listing \ref{lst:offloaddirectives}), instead a memory reference is sent from the host CPU to the micro-cores. Furthermore, the original data might not be in a memory space that is directly accessible by the device, unlike the assumptions made in GPU UVA. Whenever a micro-core reads from the variable, behind the scenes, the ePython interpreter will retrieve this value from the variable's location in the memory hierarchy, whether it be on the micro-cores or the host CPU. Likewise, if the programmer writes to such a variable then data transfer is transparently performed by ePython to where that variable is physically located. In all these cases, by default, the core will block for data transfer, either reading or writing, to complete.

Whilst it might seem that explicit fetching data from source on every access is slow, and indeed it can be, there is little choice if one is to write kernels that process large data-sets on the micro-cores. Effectively this can be thought of as a software level UM approach, but also spanning memory locations that may or may not be directly accessible by the micro-core. Driven by lessons learnt in \cite{umstreaming}, for optimisation purposes we have introduced \emph{pre-fetching}, where non-blocking data transfers are performed ahead of time with the intention that data transfer will have completed by the time the code needs to access a specific piece of data. Listing \ref{lst:prefetch} illustrates the same kernel function signature as Listing \ref{lst:offloaddirectives}, but with the programmer adding an optional \emph{prefetch} argument to the \emph{offload} decorator to pre-fetch data retrieval from the host to the micro-cores. In this case both arguments \emph{a} and \emph{b} to the \emph{mykernel} function are pre-fetched. This additional pre-fetching argument does not impact the correctness of the code, the result of computation is identical with and without pre-fetching.

\begin{lstlisting}[frame=lines,caption={Pre-fetching example by annotating the offload decorator},label={lst:prefetch}]
@offload(prefetch={a, 10, 2, 10, "readonly"}, {b, 10, 2, 10, "readonly"})
def mykernel(a, b):
  ....
\end{lstlisting}

The API signature of the \emph{prefetch} argument, is \emph{prefetch=\{variable name, buffer size, elements per pre-fetch, distance, access modifier\}} where \emph{variable name} is the name of the kernel variable argument that this pre-fetching applies to. The \emph{buffer size} argument is the number of data elements reserved in the micro-core local memory for the variable (which pre-fetching will fill up), for instance in listing \ref{lst:prefetch}, 10 integers (40 bytes) will be reserved. The \emph{elements per pre-fetch} is the number of elements to fetch on each variable access and in Listing \ref{lst:prefetch} two elements will be transferred to or from the stored data per access. The \emph{distance} argument determines when data transfer should take place, for instance in our example data will be pre-fetched 10 elements ahead. Lastly, the \emph{access modifier} argument is a further optimisation provided by the programmer which describes whether the data is mutable (potentially needs to be copied back from the micro-cores) or read only (so no copy back is required.) In the case of mutable data, we guarantee that writes complete atomically and from a single core will be performed in order. When it comes to different cores writing to the same location, whilst the atomic property is maintained there are no guarantees around the ordering constraints imposed. A by product of pre-fetching is that it retrieves multiple pieces of data (the \emph{elements per pre-fetch}) on each access which enables the overall number of data accesses is to be significantly lower than the single fetch on-demand approach, and for each of these pre-fetch requests to contain larger parcels of data. 



Passing by reference to the device, rather than eagerly copying the entire data is driven by necessity due to the limited scratch-pad fast memory on the micro-cores. A cost of pre-fetching is the memory overhead on the micro-cores where, for instance in the example of Listing \ref{lst:prefetch}, 40 bytes are required for each function argument. By making these settings explicit to the programmer they themselves can set sensible values and experiment with the most suitable settings for their application.

\subsection{Kinds for hierarchical memory}
\label{sec:kinds}
In addition to passing arguments by reference we leverage memory kinds \cite{memkind} to denote which memory space in the hierarchy variables are allocated in. A reference to this data in the specific memory space is passed to the micro-core kernel when it is invoked and data is then passed seamlessly to and from this specific location in the memory hierarchy. 

\begin{lstlisting}[frame=lines,caption={Python offload using memory kinds to control where in the hierarchy data is located},label={lst:kinds}]
from epython import offload, memkind
import random
 
nums1=memkind.Host(types.int, 1000)
nums2=memkind.Host(types.int, 1000)
 
for i in range(1000):
  nums1.append(random.randrange(0,100,1))
  nums2.append(random.randrange(0,100,1))
 
@offload
def mykernel(a, b):
  ....
  
print mykernel(nums1, nums2)
\end{lstlisting}

Listing \ref{lst:kinds} illustrates the same example code as Listing \ref{lst:offloaddirectives} but explicitly providing a level in the memory hierarchy for variables \emph{nums1} and \emph{nums2}. The API for this is found in the \emph{memkind} sub-package and this example uses the \emph{Host} memory kind initialised with the type of data it will hold (which are constants provided in the ePython module) and number of elements of this type to be allocated. We have created numerous kinds, including \emph{Host} which allocates the data in the large host memory (not accessible directly by the micro-cores), \emph{Shared} which places data in the memory which is accessible by both the host and micro-cores, and \emph{Microcore} which allocates the data in the local memory of each micro-core. Currently these kinds must reside in the host side code and are themselves are Python objects, implementing methods to copy data to and from their memory space. From the programmer's perspective, to change where in the hierarchy a variable is allocated, simply requires a single change in their code by swapping out the existing memory kind and replacing it with a different one. The underlying library and memory kind handles the low level details of this. Irrespective of where a variable is allocated, it is the reference that is passed to the micro-cores and the kinds interpret this into loads and stores. To create a kind representing a new level in the memory hierarchy requires a new Python class, inheriting from the \emph{Kind} class, with all details about that level of hierarchy encapsulated inside the kind and everything else remains unchanged.

It is still perfectly acceptable to declare variables in \emph{normal Python style} without using memory kinds, as per Listing \ref{lst:offloaddirectives}, and in such cases the variable belongs to the level of memory hierarchy that is currently in scope. Likewise, these memory kinds also abstract the \emph{declare\_on\_device}, \emph{copy\_to\_device} and \emph{copy\_from\_device} calls for managing device resident data. If a variable is allocated in the memory of the micro-cores (via the \emph{Microcore} kind), then reads and writes to these variables on the host are, under the covers, translated into copying data to or from the micro-cores using the same mechanism as the explicit calls. 






\subsection{Memory model}

Python itself does not have a standard memory model and individual implementations are free to adopt whichever memory model they wish. For instance, CPython adopts a strong memory model, relying on the global interpreter lock to enforce memory access ordering. In contrast ePython adopts a weaker memory model to optimise multi-core performance.

Whether it be the eager or pre-fetching of data, whenever a micro-core attempts to accesses a scalar variable or index of an array, held elsewhere in the memory hierarchy, preference is given to any local copy held on that micro-core. If there is no local copy, then a data transfer will be performed. For instance, with the statement \emph{a = a * a}, ePython will check whether \emph{a} is held locally and if not will retrieve the corresponding data. This local copy will then be used for all the reads (i.e. \emph{a * a}) and the write occurs both to the local copy of \emph{a} and is also written back to the variable's location on the host. Due to memory limits of the micro-cores, it might be that locally held copies of data elsewhere in the memory hierarchy are freed. This is especially the case with the eager fetching approach which, unlike pre-fetching, does not allocate any user defined buffer space and instead uses a central storage pool. Access to data, whether it be a scalar or array element, held in memory locations outside the core will always first check whether there is a copy held locally, and if not perform the explicit data movement required.

For the two statements \emph{tmp = a; a = tmp * a}, on each access of \emph{a} ePython will check whether a copy of the data is held locally, and if not perform necessary data movement. Based on these two statements side by side, it is highly likely that the copy of \emph{a} from the first statement will still be resident for the access in the second statement. The write of \emph{a} in this example will update both the local copy and also the variable held in the memory hierarchy.

Therefore, within a core, updates to data are in-order and atomic. Between the cores the model is weaker for performance reasons and the ability to reuse data held locally rather than explicitly fetch each time. This provides a simple and consistent model, requiring limited support from the hardware and runtime software. The programmer should be aware of this because, if two kernels are working with the same data and both reading and writing to this, then ePython only imposes the atomicity of these updates. There is no guarantee around the order in which accesses from different cores will complete, or when kernels will see the data written by kernels on other cores. This is a somewhat different than that adopted by many multi-core CPUs, which tend to only write data on cache flush but do support a stronger memory model, often via directory based cache coherence.

\section{Implementation}
\label{sec:implementation}

As outlined in Section \ref{sec:directives}, passing kernel arguments by reference and the addition of memory kinds extends the approach of offloading kernels and interacting with device resident data. These changes not only impact the behaviour of the language, but also require extensions to the ePython interpreter. The purpose of this section is not to examine all the low level changes required, but instead provide a high level view of how we implemented these new features as we believe this is also applicable to other dynamic languages. Given the very limited on-core memory, adding support for pass by reference and memory kinds resulted in a significant challenge as the approach had to be both usable by the programmer and also implementable given the memory constraints.

\begin{figure}
\centering\includegraphics{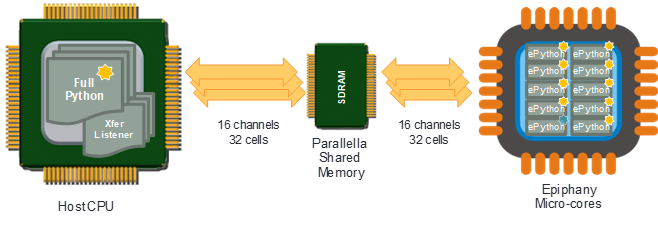}\caption{ePython reference communications architecture}

\label{fig:xref}
\end{figure}

The first step in supporting this new behaviour was the underlying data transfer code, connecting the host CPU with the micro-cores. Figure \ref{fig:xref} illustrates our approach where the host's shared memory is used to provide a direct link between Python running on the CPU and the ePython VM on each micro-core. A number of \emph{channels} are constructed, one per core, and each channel contains thirty two 1KB \emph{cells}. This enables up to thirty two concurrent transfers between the host CPU and each micro-core.


Each ePython interpreter running on a micro-core maintains it's own symbol table which, for each variable, contains some metadata and a pointer to the physical data in either the stack or heap. We extended the symbol table metadata to add an extra \emph{external} flag indicating whether the pointer references directly accessible or external, non-directly accessible, data. When passing kernel arguments by reference to the micro-cores, the variable's \emph{external flag} in the symbol table and pointer to this reference are set appropriately. Whenever Python code running on the micro-cores accesses a variable it will first check this external flag in the symbol table. If the flag is zero then a direct access is issued as per normal, but if it is one then the corresponding data is held externally and extra calls are required by the interpreter.

Extra calls for interacting with external data have been added to the ePython runtime, which the interpreter calls when external access is required. These additional functions can be thought of as blocking and non-blocking primitive data communication calls, which the programmer themselves never sees. The blocking calls, to copy data on or off the device are the simplest, and code execution on the micro-core will block until data access has completed. Pre-fetching requires non-blocking data transfer calls, where the core will request data ahead of time, continue working and then have some way of tracking whether the access has completed when the data is required. Non-blocking external data access functions, again in the runtime, return a handle which corresponds to a specific data transfer cell in the micro-core's channel. A \emph{ready} function is provided by the runtime to test for completion. 



The modifications described here; changes to the symbol table and extra runtime support are the only extensions required in the ePython interpreter and runtime running on the micro-cores. All other aspects of our abstraction are resident on the host CPU, which is not memory constrained, and effectively translate into these lower level primitives. This is important due to the memory limits of the micro-cores and the extensions discussed here require an extra 1.2KB of memory on the micro-core for the interpreter and runtime. Bearing in mind this enables the programmer to, for the first time in ePython, work with arbitrarily large amounts of data held anywhere in the memory hierarchy we believe it is a price worth paying.

The host CPU side must be able to identify what each reference corresponds to, and then decode this and perform physical memory access. In reality, the reference itself isn't a physical memory location but instead a unique identifier which is used to look up the corresponding variable and memory kind it belongs to. This information is then passed to the associated memory kind which \emph{decodes} the reference and performs appropriate action(s). Lookup on the host side has been designed this way for further extensibility, where the memory kinds could perform some functionality other than memory access, such as communicating with remote memory spaces or IO.


\section{Results and evaluation}
\label{sec:results}
The data science bowl \cite{datasciencebowl} is a prominent data science competition with significant social impact. In 2017 the challenge was held around the development of lung cancer detection algorithms, with the National Cancer Institute (NCI) making available thousands of high-resolution 3D lung scans. The aim is to develop techniques and approaches for determining whether lesions are cancerous or not, as the current generation of detection technology is plagued with false positives. This paper is using the NCI's data differently to the competition and asking a separate question. Instead of being concerned with the absolute accuracy of prediction, we are instead evaluating whether micro-core architectures and the parallelism that they provide can benefit the area of machine learning. Accuracy of detection is the primary concern for the competition, but the execution of these algorithms also needs to be realistic. Not only does this involve training the model in a timely fashion, but also employing an architecture which is affordable and utilises a minimal amount of power, which is where micro-core architectures are of main interest. 

In \cite{epython-poster} we developed a simple neural network with one hidden layer of 100 neurons which splits the 3D CT lung scans into two groups, 70\% for training and 30\% for testing. In this approach the input pixels of the image are distributed amongst the micro-cores which are used for accelerating the linear algebra involved in training and model and the back-prop. Parallelism comes from the fact that each micro-core is operating on a separate part of the overall image and previously each image was copied on to the micro-cores on kernel invocation. Our new offload behaviour mean that these images now remain in host memory and instead a reference to them is passed to the micro-cores on kernel invocation. Our previous eager copy approach was shown to perform competitively against Python and native implementations, but the limited memory of micro-cores meant that images had to be interpolated down from a maximum on-disk size of over 100MB to a size that the input data and neural network could fit within the shared chunk of main memory (e.g. 32MB on the Epiphany/Parallella configuration). In this paper we are using this same code as a benchmark but crucially the modified behaviour of kernel invocation as described in Section \ref{sec:directives} means that we can run the full sized images for the first time. In our opinion this moves micro-cores and ePython from being an interesting research technology, to becoming more mature and a more serious contender for these real world applications. 


\subsection{Experimentation results}
\label{sec:actualresults}
\begin{figure}[h!]
\centering\includegraphics[width=0.8\textwidth]{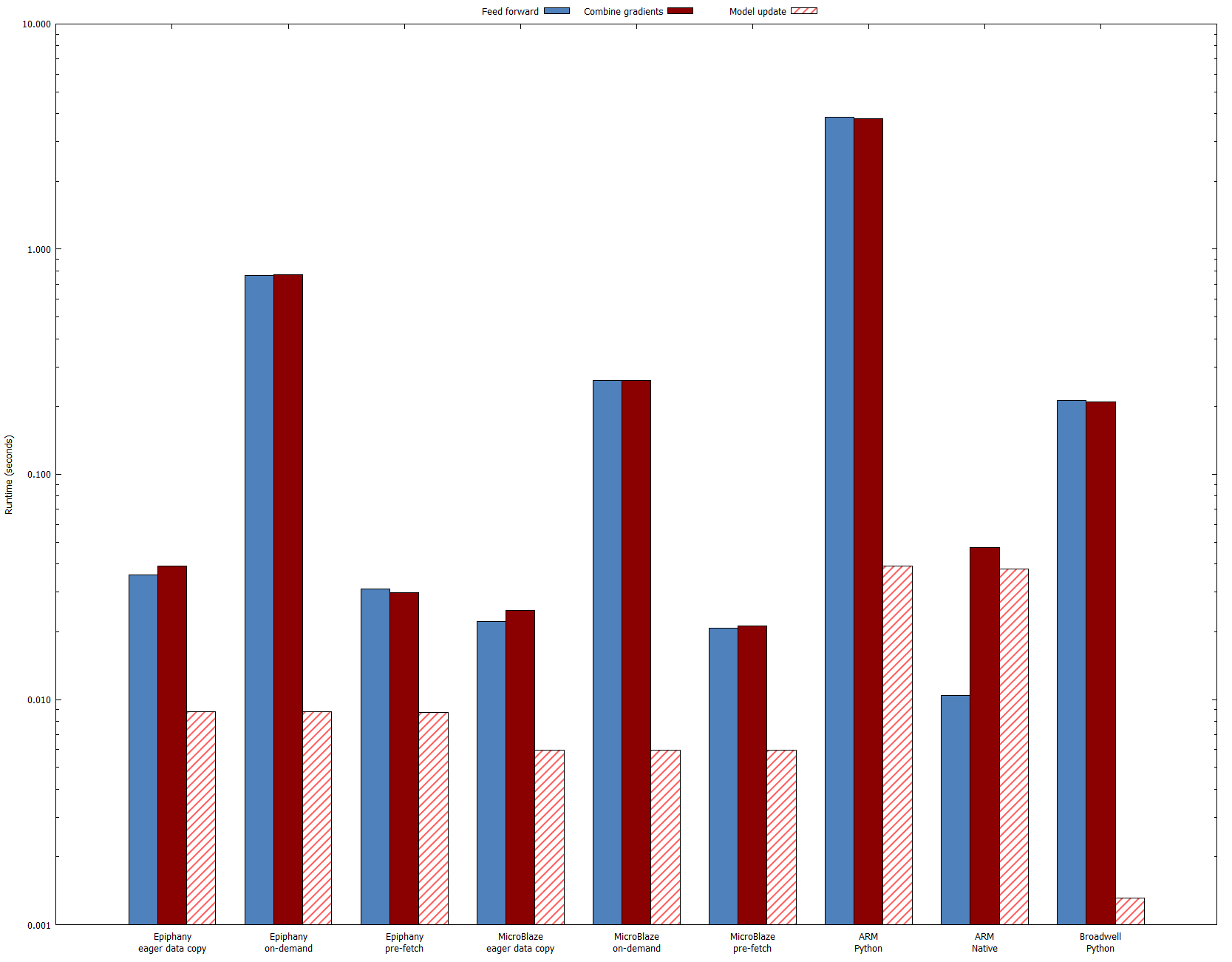}\caption{Machine learning benchmark results for small, interpolated, images}

\label{fig:mlresultssmall}
\end{figure}

Figure \ref{fig:mlresultssmall} illustrates performance results for ePython with our new offload behaviour for both on-demand and pre-fetching, against the previous eager data copying on kernel invocation under ePython. Also included are runs on the ARM host using CPython for the kernels and a native implementation which calls into Numpy for the kernels, which has been compiled with GCC at optimistion level 3. There is also an implementation via CPython on Broadwell, where each ARM and Broadwell result is based on execution on a single core. For each configuration, there are three results; \emph{feed forward} is the time taken to do a forward pass of the neural network, \emph{combine gradients} is the time taken to calculate gradients for a batch of training data and \emph{model update} is the time taken to update the model with gradients for the batch.

The results in Figure \ref{fig:mlresultssmall} represent the scaled down, interpolated, images as per experiments in \cite{epython-poster} running on both the Epiphany and MicroBlaze. For these experiments we have 3600 input pixels distributed amongst the micro-cores, with a hidden layer of 100 neurons. There are two key data structures, a matrix of input-hidden layer weights distributed among the micro-cores and a vector of hidden layer-output neuron weights. Each small image, passed for kernel invocations is 14.4KB. Forward feed involves a dot product on the weight matrix with the image, and a second dot product on the resulting values with the hidden layer-output neuron weight vector. Combining gradients, done for each image (but we don't update the model weights until after the batch) involves a dot product and an outer product. For these small images each kernel involves around 45000 floating point operations.

It can be seen that the original ePython kernel invocation version, \emph{ePython eager data copy}, compares favourably against CPython for both the Epiphany and MicroBlaze, and native versions on the ARM host which is due to the parallelism provided by the Epiphany. The \emph{ePython on-demand} versions represent the benchmark relying on the modified behaviour described in Section \ref{sec:directives}, with accesses done on-demand and not taking advantage of pre-fetching. The \emph{ePython pre-fetch} results represent a version of the code using our modified behaviour and pre-fetching optimisation. 

For both the Epiphany and MicroBlaze, the \emph{on-demand} version of this benchmark is significantly slower than the existing, eager data copy, behaviour of ePython. This is because the micro-cores retrieve individual elements of data, one at a time, and for each of these it must block until the transfer has completed. In contrast the pre-fetch version of the benchmark provides up to 1.3 times better performance for the calculation of gradients on the Epiphany than the existing eager data copy ePython implementation and is around 25 times faster than the on-demand data copy approach for the Epiphany. The pattern is similar for the MicroBlaze, although the differences are less. There is no change in the model update runtimes because this does not rely on data transfer. The performance improvement of pre-fetching over eager data copying is due to two factors, firstly the kernel can start as soon as the single reference is copied across rather than the entire data, and secondly our new data transfer mechanism enables the ePython module running in CPython to communicate directly with the ePython VM on the micro-cores rather than having to go via the ePython host process.


\begin{figure}[h!]
\centering\includegraphics[width=0.8\textwidth]{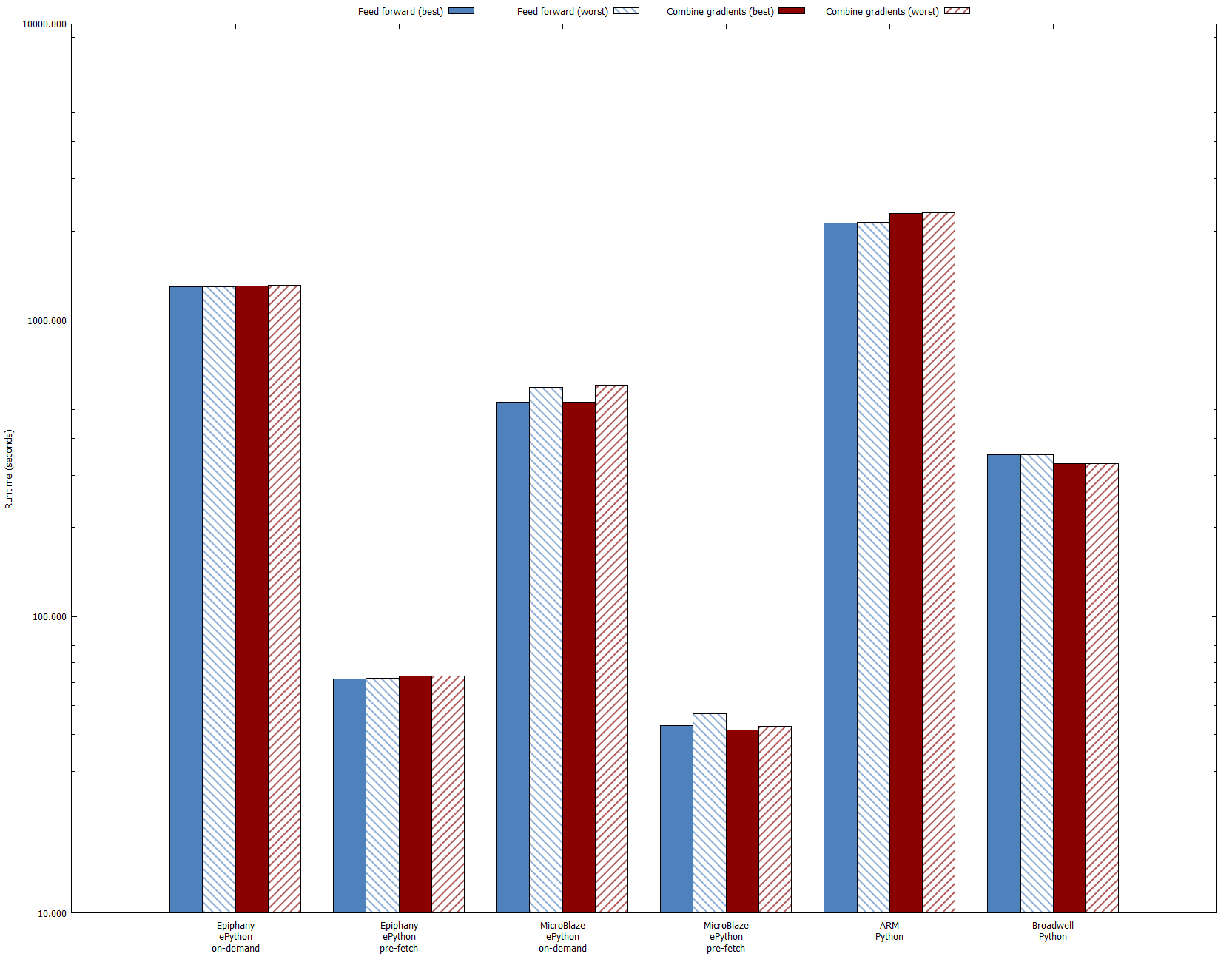}
\caption{ePython machine learning benchmark results (full sized images)}

\label{fig:mlresultslarge}
\end{figure}

Before the work of this paper it was impossible for these Python kernels to process the full sized images on the micro-cores. Figure \ref{fig:mlresultslarge} illustrates the performance of the machine learning benchmark for a forward pass through the neural network (\emph{feed forward}) and calculation of gradients (\emph{combine gradients}) on the Epiphany and MicroBlaze micro-cores when running with the full sized images (again both \emph{on-demand} and \emph{pre-fetch} versions) and CPython on the ARM host using the same sized hidden layer as previously. Similarly to the small images, enabling the pre-fetching of data is much faster than the on-demand approach, especially for the Epiphany where it is around 21 times faster. The full sized images are, on average, around 7 million pixels which is 1966 times larger than the small, interpolated, images of 3600 pixels. The average, single precision, input data that must be transferred to the micro-cores for each kernel is around 30 MB.

\begin{table}[h!]
\centering%
\footnotesize
\begin{tabular}{| c c c c |}
\hline
\textbf{Technology} & \textbf{MFLOPs} & \textbf{Watts} & \textbf{GFLOPs/Watt} \\
\hline
\textbf{Epiphany III} & 1508.16 & 0.90 & 1.676 \\
\textbf{MicroBlaze} & 0.96 & 0.19 & 0.005 \\
\textbf{MicroBlaze + FPU} & 47.20 & 0.18 & 0.262 \\
\textbf{Cortex A-9} & 33.20 & 0.60 & 0.055 \\
\hline
\end{tabular}\caption{Performance and power consumption for LINPACK benchmark}
\label{tbl:energyml}
\end{table}

ePython is an interpreter, therefore to explore performance and power efficiency in more detail, and avoid noise due to the interpreted nature of ePython, we modified the C LINPACK benchmark \cite{dongarra2003linpack} to run on the micro-cores. We measured the voltage and amperage of the board using two UNI-T UT60E multimeters and this power usage, along with the benchmark's measured floating point performance (in MFLOPs) for both technologies, plus an embedded class ARM Cortex-A9 for comparison, are detailed in Table \ref{tbl:energyml}. We have included results for the integer-only and hardware floating point (FPU) MicroBlaze soft-cores to highlight the significant impact of software floating point emulation on kernels such as LINPACK. It can be seen that the Epiphany provides a much greater FLOP rate, 31 times, that of the MicroBlaze with FPU. This is, in part, due to fact that the Epiphany contains sixteen cores running at 600Mhz, verses eight MicroBlaze cores at 100Mhz. If we normalise the core count and clock rates, the Epiphany is still about 3 times faster per core than the MicroBlaze with FPU.

For comparison, a recent study \cite{castro2018energy} illustrated that performance per Watt on the Pascal GPU is 42 GFLOPs/Watt and on the Maxwell 23 GFLOPs/Watt for a similar machine learning problem. Whilst these results are significantly higher than those achieved in the micro-core LINPACK benchmark, crucially these two HPC grade GPUs draw a maximum of 250 Watts, whereas the power draw of the micro-cores used in our experiments was 0.90 Watts for the Epiphany and 0.18 Watts for the MicroBlaze. This much smaller power draw is very important because it means that they are highly applicable to high performance embedded devices, where absolute power draw is very important. Bearing in mind that the Pascal is a smaller process size, 16nm, than both the Epiphany and Zynq-7020 used here, this newer hardware will inevitably exhibit some power benefits. When one considers that the Maxwell is based on 28nm technology and the Epiphany on 68nm, the performance per Watt differences between these two technologies becomes more understandable. The Zynq-7020 FPGA alone has a theoretical performance per Watt of 72 GFLOPs/Watt \cite{fpgavsgpu}, and even though achieving the theoretical peak is not realistic, our results for the LINPACK benchmark indicate that the use of soft-cores sacrifices this power efficiency significantly, which aligns with the work done in \cite{lysecky2005study}. It is our feeling that the performance of the soft-cores, not least the fact that they are running at 100Mhz in our experiments, is the main limit here and if this can be increased, for instance by using more advanced FPGAs, redesigning the block layout or by customising the soft-core designs for specific applications \cite{kadi2018general} using techniques such as ‘warp’ processors \cite{lysecky2004warp}, then this will benefit the performance per Watt metric too.

Another recent study \cite{embedded-energy} considered performance and power efficiency of other technologies more commonly found in embedded computing and hence closer to micro-cores. They used a benchmark based on LINPACK and NVIDIA's Jetson TX1 (with a Tegra X1 GPU) achieved 16 GFLOPs, drawing a maximum of 15.3 Watt and performance per Watt of 1.2 GFLOPs/Watt. ARM's quad core Cortex A53 CPU achieved 4.43 GFLOPs, drawing a maximum of 5.1 Watts and achieving 1.07 GFLOPs/Watt. For comparison, a sixteen core Haswell CPU achieved 47.7 GFLOPs, drawing 29.1 Watts and delivering 1.64 GFLOPs/Watt. When considered against the results for these technologies, especially the Jetson TX1 and ARM designed for embedded systems, the results obtained for the Epiphany and MicroBlaze are more respectable. 

When comparing power consumption (Watts) of the micro-cores running LINPACK, we find that the Epiphany requires five times as much power as the MicroBlaze with FPU, and about twice the power of the ARM Cortex A-9. However, when we consider the power consumption / performance ratio (GFLOPs/Watt), the situation is reversed, with the Epiphany being about 6 times more efficient than the 8-core MicroBlaze and about 30 times more efficient than the Cortex-A9. The micro-core LINPACK benchmark results are only impacted by the time it takes for the device kernels to respond to execution requests and the acknowledgement of completion. Therefore, the results in Table 1 are not impacted by communications link bandwidth restrictions. This is why we see a greater difference in results between the Epiphany and MicroBlaze for LINPACK than for the Machine Learning benchmark, where there is a significant amount of data transfer. Quite simply, the bandwidth to the Epiphany chip is significantly less than the Zynq-7020 and this explains why, in Figures \ref{fig:mlresultssmall} and \ref{fig:mlresultslarge}, even though the MicroBlaze's computational performance is far more limited due to the lower clock rate, the performance it delivers is still competitive with the Epiphany. This also explains why, in Figure \ref{fig:mlresultslarge}, the on-demand version for the Epiphany is so much slower than the MicroBlaze version. From experimentation we found that, on the Epiphany/Parallella configuration the maximum bandwidth we could get with our benchmark was 88 MB/s but this frequently dropped to as low as 16 MB/s (theoretical peak is 150MB/s), whereas on the MicroBlaze/Pynq-II we consistently achieved around 100 MB/s (theoretical peak is 131 MB/s).

Whether it be running on the Epiphany or the MicroBlaze, there is a significant performance difference between the \emph{on-demand} and \emph{pre-fetch} approaches. To help understand the reasons behind this, and also explore how the size of data transfer impacts the overall load time, a synthetic benchmark was written to accurately measure the message load time on the micro-cores. This benchmark measures the time that the micro-core is stalled whilst data is copied from the host onto the micro-core. 





\begin{table}[h!]
\centering%
\footnotesize
\begin{tabular}{| c c c c c c c |}
\hline
& \textbf{\makecell{128B \\ on-demand}} & \textbf{\makecell{128B \\ pre-fetch}} & \textbf{\makecell{1KB \\ on-demand}} & \textbf{\makecell{1KB \\ pre-fetch}} & 
\textbf{\makecell{8KB \\ on-demand}} & \textbf{\makecell{8KB \\ pre-fetch}} \tabularnewline \hline
\textbf{Min} & 0.099 & 0.098 &  0.759 & 0.758 & 6.396  & 7.215\tabularnewline
\textbf{Max} & 0.112 & 0.111 & 0.955 & 0.913 & 11.801  & 9.452\tabularnewline
\textbf{Mean} & 0.104 & 0.103 & 0.816 & 0.804 & 7.882  & 8.537\tabularnewline
\hline
\end{tabular}\caption{Synthetic benchmark micro-core stall time for different data sizes (msecs)}
\label{tbl:loadtimes}
\end{table}

Table \ref{tbl:loadtimes} illustrates the results from this benchmark, with minimum (best case), maximum (worst case) and mean timings against the data size and access configuration. The reader can imagine the data size here representing one (very large) element of data for the on-demand approach and for pre-fetching it represents the size of the chunk of data (the \emph{elements per pre-fetch} of Section \ref{sec:passbyref}) retrieved on each access. The major reason for variation in timings for a specific configuration (the minimum and maximum) is that a dedicated thread on the host CPU needs to pick up a request and handle it, with other activities on the same CPU this response time can vary. 

Until 8KB, the average and maximum load time for the on-demand approach is higher than that of the pre-fetched approach. At the largest data size of 8KB, the maximum time is still largest for on-demand but the mean time is lower for on-demand in contrast to pre-fetched, as is the minimum time. Given the performance results for the machine learning benchmark this was unexpected and is due to the extra overhead of pre-fetching. Because of the size of the data, transfer takes longer, and the core runs out of work and-so must block for much of the time. The more complex pre-fetch protocol, where the interpreter continually calls into the \emph{ready} function of the runtime to check for data, adds much of this additional overhead when compared to the on-demand approach where the is blocking behaviour is simpler.

This benchmark only measures stall time for a single load, and it can be seen that there is only a small difference between on-demand and pre-fetching, not the 21 to 25 times difference that was seen with the machine learning benchmark. Instead, the reason for poor performance of the on-demand machine learning benchmark was that this makes individual requests for each element of data which swamps the communication channels and keeps the host CPU very busy, continually responding to these requests. In contrast, the pre-fetch approach retrieves data in chunks and the fact that there are significantly fewer requests made is most important for performance. In terms of the optimal data transfer size (\emph{elements per pre-fetch} specified by the programmer for pre-fetching) this depends heavily on the application. For the benchmark, load time is significantly less for smaller data sizes, but for a real world application larger data sizes will reduce the number of requests the host must service. This will likely be especially important if the host is also required to run some part of the code whilst the micro-cores are active.

\subsection{Programmability concerns}
So far we have considered the performance of our benchmark on the Epiphany and MicroBlaze micro-core architectures. The programmability of this code should also be considered, and specifically the qualitative differences writing code in other programming technologies that target these micro-cores. For both these architectures, the most obvious approach would be to use C and interact with the shared memory directly. From a programming perspective this is a very significant challenge because, whilst these cores can directly access some part of the host shared memory, similar to NVIDIA UVA, placing the data in this shared location and using it directly incurs significant performance penalty due to limited off chip bandwidth as we have shown in Section \ref{sec:actualresults}. Therefore, to get good performance necessitates the programmer writes explicit code to perform data copying to core local memory and piping it in ahead of time. Further challenges here are the weak hardware memory model of the micro-cores and on the Epiphany, only 32MB of main memory is directly accessible to the micro-core which even a single, full sized image, does not fit into. This is all possible to be developed, but requires significant programming expertise, is often bespoke to a specific code, error prone, and involves the programmer spending significant time on the tricky, low level aspects rather than their application logic. 

Whilst other higher level approaches, such as OpenCL, have been developed for the Epiphany, this still requires fairly significant modifications on the host side and the full range of OpenCL features is unavailable due to the limited on-device memory. Instead, using our approach, the programmer is far more abstracted from the mechanics of how data transfer occurs, more like a software implementation of NVIDIA's UM but with more control over exactly where in the memory hierarchy we are located using memory kinds, with minimal code level changes required. It is also important to highlight that the programmer has significant flexibility to experiment with concerns of data placement where, just by changing the memory kind, they can move the location of their data with the kinds themselves handling how this happens.

\section{Conclusions}
\label{sec:conclusions}
In this paper we have described the abstractions required for enabling micro-core architectures to handle arbitrary large amounts of data held in different memory spaces. By changing the behaviour of kernel invocation to a pass by reference model, and combining this with memory kinds, the programmer can manage their data locality and movement whilst still being abstracted from the lower level details. With a single change in memory kind the programmer is able to trivially experiment with placing data in different levels of the memory hierarchy.

Using a machine learning benchmark, we have demonstrated that these concepts open up the possibility of running kernels on micro-cores with arbitrarily large data sets and the more constrained the off-chip bandwidth, in the case of the Epiphany/Parallella combination, the more important the pre-fetching optimisation becomes. Whilst we have used Python, the Epiphany, and the MicroBlaze as vehicles for developing and testing our ideas, crucially the work described in this paper is not just limited to these technologies and defines offloading semantics for the entire class of micro-core architectures which are becoming more and more widespread. As we argued in Section \ref{sec:otheroffload}, the design of micro-cores is fundamentally different from that of other accelerators and as such how programming technologies offload kernels, and specifically deal with data, must be handled differently.

We have seen that, when moving from an eager copy to pass by reference model, it is important to use pre-fetching in order to obtain best performance. Interestingly a significant aspect of pre-fetching, in addition to data transfer occurring whilst the core is busy and hence avoid the micro-core stalling, is that pre-fetching retrieves data in chunks rather than single individual elements that the default on-demand approach requires, which is broadly in line with experiences of UVA and UM memory movement strategies as described in \cite{umstreaming} and \cite{landaverde2014investigation}. 


We have considered power efficiency, with the  Epiphany delivering up to 1.7 GFLOPs/Watt and the MicroBlaze 0.262 GFLOPs/Watt. Whilst power efficiency is competitive with technologies designed for the embedded space, both in terms of performance per Watt and power draw, when compared against latest generation HPC grade GPUs we can see that these other technologies provide much greater performance per Watt at the cost of significantly higher overall power draw. Generation wise, the closest comparison is the Maxwell GPU against the Zynq-7020 running our MicroBlazes, as these are both 28nm technology. It is our feeling that the MicroBlaze soft-core on the Zynq-7020 is significantly under-performing, and whilst it is beyond the scope this being one of our focus on hierarchical memories, it would be a very interesting to explore alternative soft-cores such as RISC-V to understand whether this same behaviour holds. 

The reader might wonder why we did not make the pre-fetching optimisation the default option and indeed the results of this paper indicate that it would be sensible to do so. However pre-fetching can be difficult to do correctly as it adds memory transfer and instruction level overhead \cite{hadade2018software}. Indeed \cite{hadade2018software} argues that auto-tuning for CPU cache pre-fetching is crucially important and, we believe going forwards a similar auto tuning approach would be useful here. Especially as our optimal pre-fetching arguments, which were found empirically, were different between large and small image benchmark runs, and micro-core technologies. 




Whilst this paper has focused on micro-cores we also believe that the work here has a wider applicability. The recently announced European Processor Initiative will combine many ARM cores together on the CPU and utilise the RISC-V architecture as a basis for accelerators \cite{epi}. There are numerous soft-core RISC-V implementations, many of which follow a similar pattern of large numbers of cores each with small amounts of memory. OpenMP has been suggested as an approach to programming such a future machine and the ideas discussed in this paper will likely be applicable.


\bibliographystyle{model1-num-names}
\bibliography{sample.bib}







\end{document}